\documentclass{PoS}
\pdfoutput=1
\usepackage{amsmath}
\usepackage[section]{placeins}
\usepackage{slashed}
\usepackage{url}
\title{Exploring the phase structure of 12-flavor $SU(3)$}
\ShortTitle{Phase structure of $SU(3)$ with $N_f = 12$}
\author{\speaker{Zechariah Gelzer}\\
        Department of Physics and Astronomy, University of Iowa, Iowa City,
        Iowa, USA\\
        E-mail: \email{zechariah-gelzer@uiowa.edu}}
\author{Yuzhi Liu\\
        Department of Physics, University of Colorado, Boulder, Colorado, USA\\
        E-mail: \email{yuzhi.liu@colorado.edu}}
\author{Yannick Meurice\\
        Department of Physics and Astronomy, University of Iowa, Iowa City,
        Iowa, USA\\
        E-mail: \email{yannick-meurice@uiowa.edu}}
\abstract{We are studying the SU(3) gauge theory with 12 staggered fermions,
          searching for the endpoint of the line of first-order phase
          transitions in the mass--beta plane. This endpoint plays an important
          role in our understanding of the phase diagram of this model. Having
          found this endpoint with high statistics on a small lattice using
          unimproved staggered fermions, we are working to find it on larger
          lattices and with improved actions. For an action improved with
          nHYP-smeared staggered fermions, we discuss the effect of slowly
          turning off the improvement on the broken shift symmetry phase.}
\FullConference{The 32nd International Symposium on Lattice Field Theory\\
                23-28 June, 2014\\
                Columbia University New York, NY}
\begin{document}

\section{Motivations}

Following the recent discovery of a Higgs-like boson with an approximate mass of
126 GeV, there is a possibility that it exists as a composite particle formed by
a new strongly-interacting gauge theory. To study the theory candidates, we must
better understand their phase structures, such as whether a given model develops
an infrared fixed point (IRFP) \cite{DeGrand:2011ptrsa}. Models that have been
receiving increasing attention include $SU(3)$ gauge theories with $N_f = 8,12$
flavors \cite{Jin:2009posl, Fodor:2012ahl, LatKMI:2013prl}.

We are studying $SU(3)$ gauge theory with 12 staggered fermions, adopting both
unimproved and improved actions. We perform simulations by using the standard
hybrid Monte Carlo (HMC) algorithm, applying code by Dr.\ Donald Sinclair and
Dr.\ Yuzhi Liu for the unimproved case, as well as a modified version of the
MILC code \cite{MILC} for the improved case. We pay particular attention to the
line of first-order phase transitions in the $m$--$\beta$ plane, where $m$ is
our fermion mass and $\beta$ is inversely proportional to the bare gauge
coupling ($\beta = 6 / g^2$ in $SU(3)$). This question of an endpoint for
$SU(3)$ with $N_f = 12$ is also being discussed by groups using a gauge action
with renormalization group (RG) improvement, known as doubly blocked Wilson
(DBW2) \cite{Jin:2013ahl}.

The existence and location of the $m$--$\beta$ transition endpoint will help us
determine important features of the theory's phase diagram, primarily since all
RG flows start from the $m$--$\beta$ plane \cite{Liu:2013posl}. One may then
analyze the endpoint from the viewpoint of Fisher zeros: zeros of the partition
function $Z$ in the complex $\beta$-plane, at which phase transitions occur.
Such Fisher zeros act as indicators of an IRFP for RG flows
\cite{DenBleyker:2010prl} and are thus vital to understanding the continuum
behavior of lattice models.

\section{Numerical results}

\begin{figure}
    \begin{center}
    \includegraphics[width=0.60\textwidth]{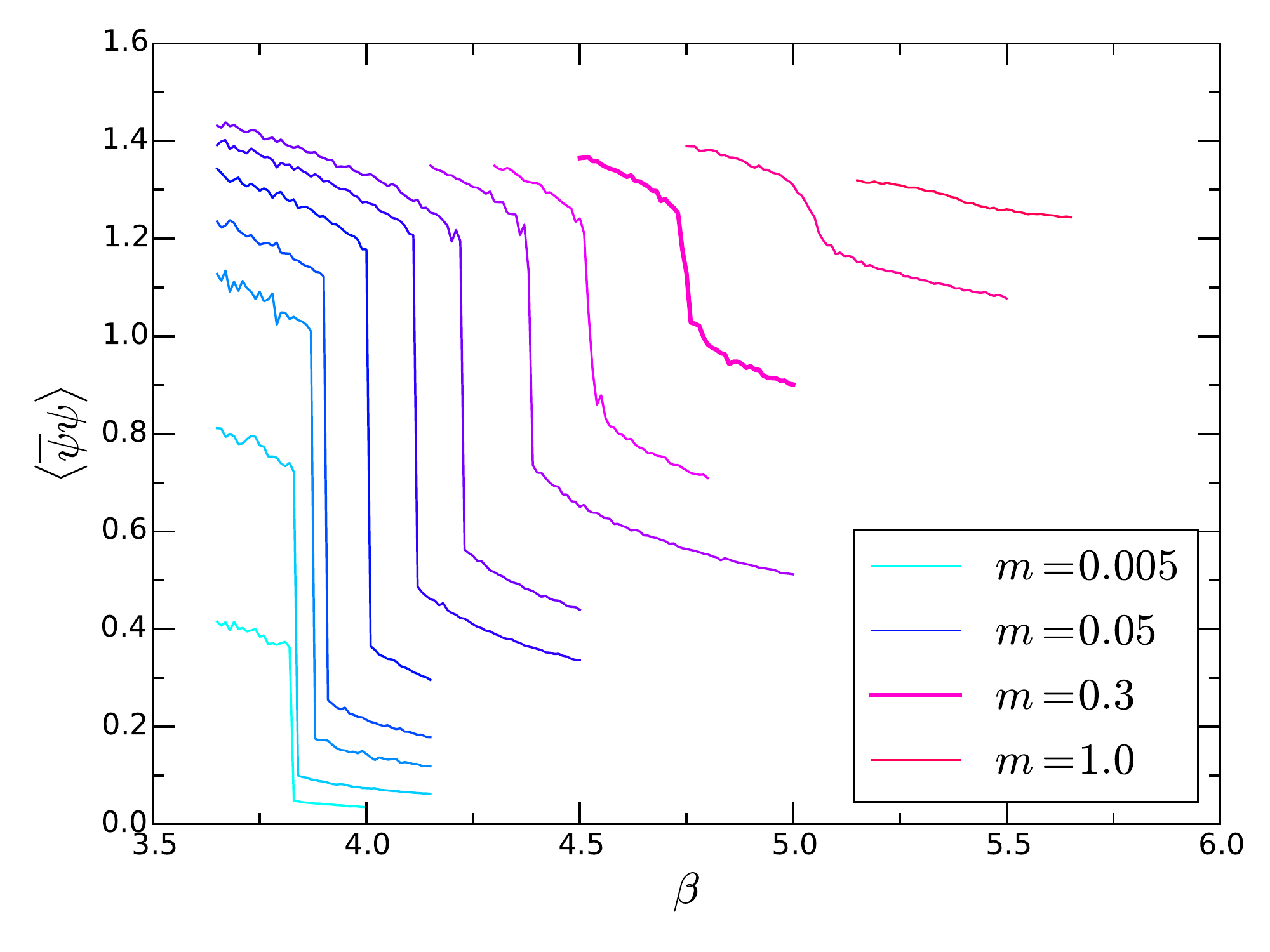}
    \caption{Unimproved HMC.
             Chiral condensate vs.\ $\beta$ for increasing $m$, with $N_f = 12,
             ~V=4^4$. The masses included (from left to right) are as follows:
             0.0050, 0.0105, 0.0200, 0.0300, 0.0500, 0.0755, 0.0995, 0.1505,
             0.2002, 0.3000, 0.5000, 0.9999.}
    \label{fig:rhmc_l44_pbp}
    \end{center}
\end{figure}

\begin{figure}
    \begin{center}
    \includegraphics[width=0.45\textwidth]{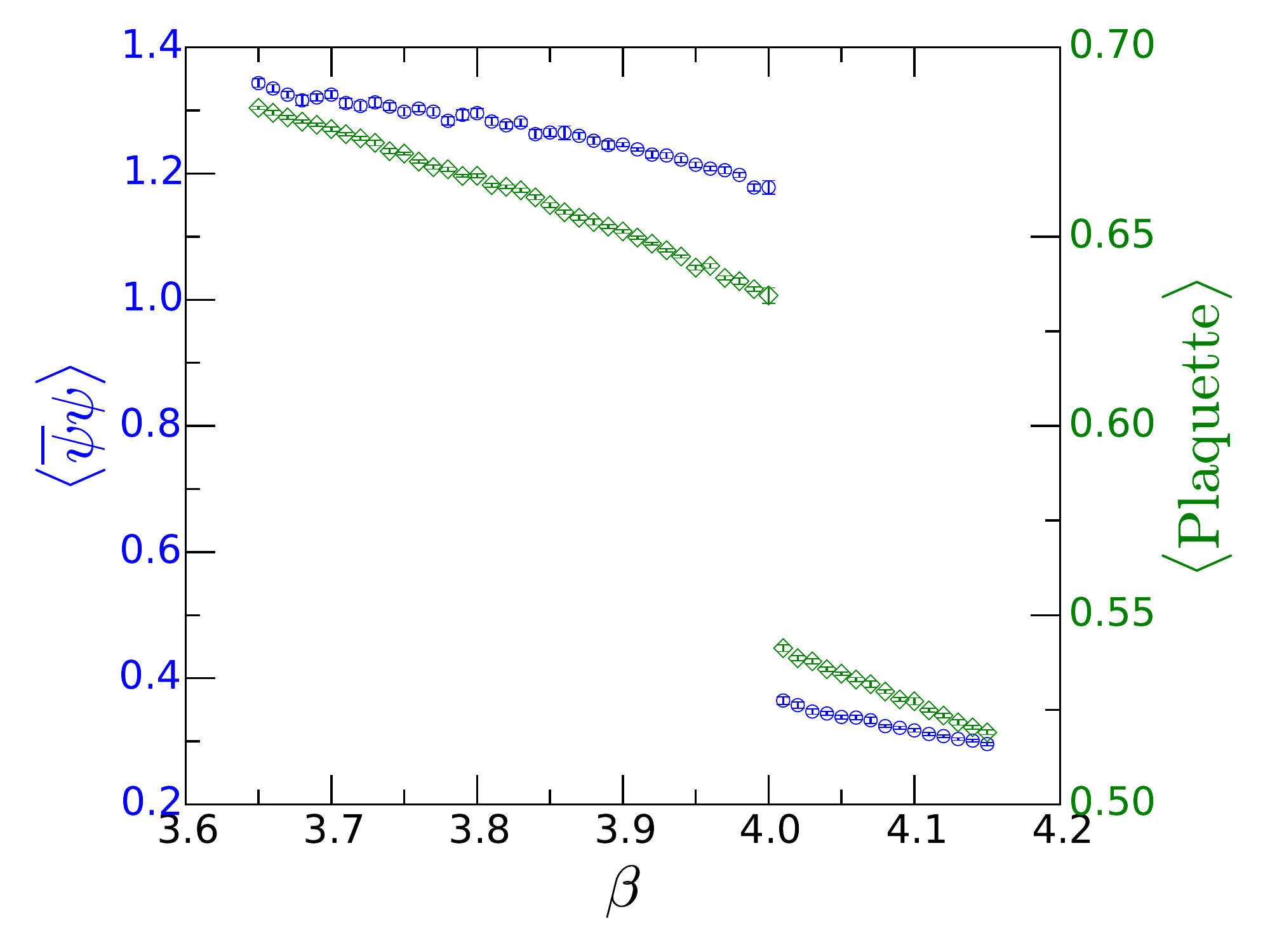}
    \hspace{0.05\textwidth}
    \includegraphics[width=0.45\textwidth]{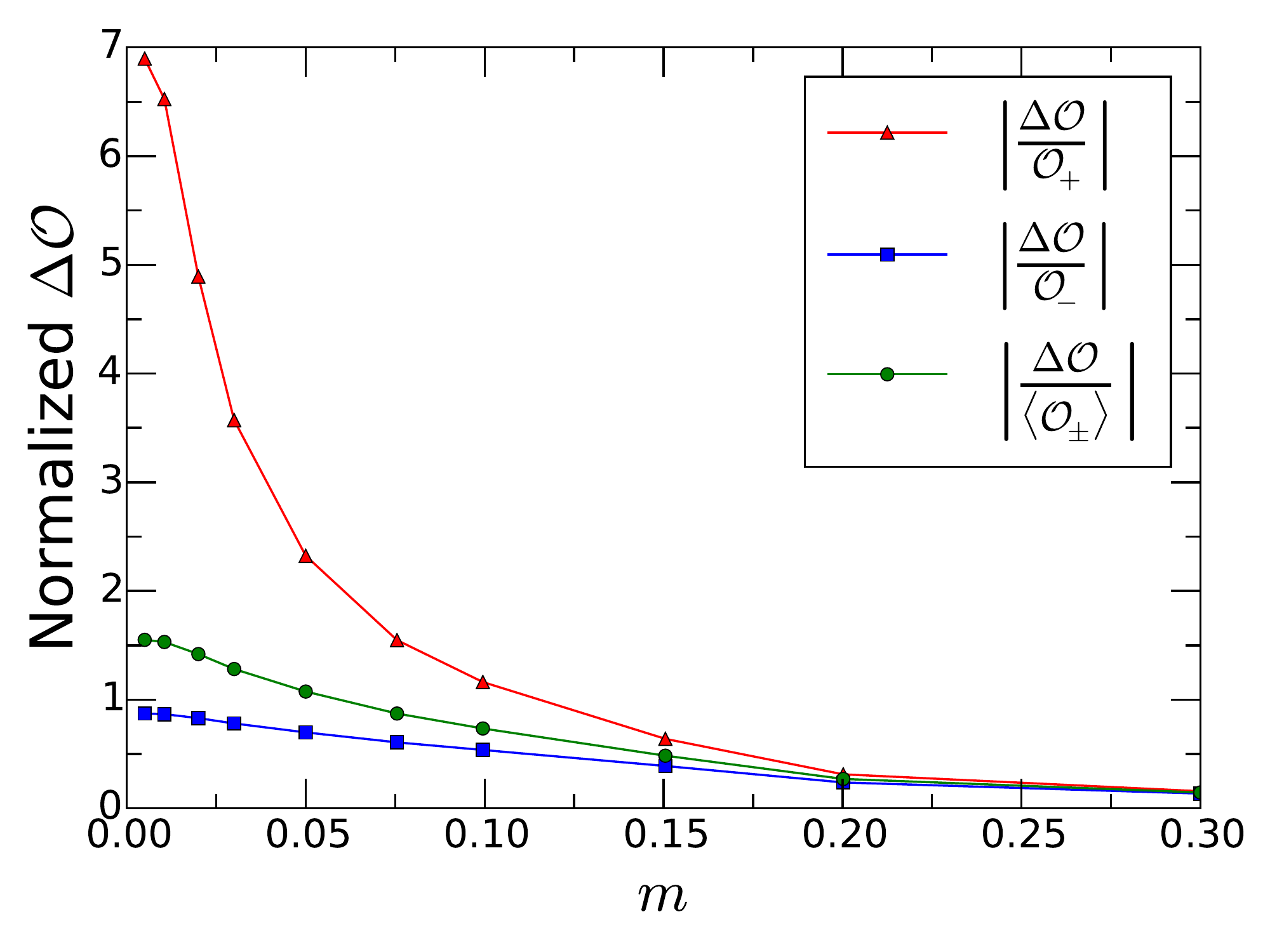}
    \caption{Unimproved HMC.
             Left panel: chiral condensate (blue) and average plaquette (green)
             vs.\ $\beta$, with $N_f = 12, ~V=4^4, ~m=0.05$. Note that the error
             bars are smaller than the data point markers.
             Right panel: normalized transitions $\Delta \mathcal{O}$ about the
             critical beta $\beta_C$ for $\mathcal{O} = \langle \overline{\psi}
             \psi \rangle$ vanish with increasing $m$. The red curve is
             normalized to $\beta > \beta_C$, the blue one to $\beta < \beta_C$,
             and the green one to an average of the two regions.}
    \label{fig:rhmc_l44_errors-trans}
    \end{center}
\end{figure}

First, we employ unimproved staggered fermions and the simplest form of the
Wilson gauge action, wherein we are limited to plaquette terms in the
fundamental representation only (i.e., $\beta_A = 0$). Starting from the HMC
algorithm, which has been made exact by its global Metropolis accept/reject
step, we simulate with high statistics on small lattices to quickly yield
observables with small errors. Completing simulations for a range of fermion
masses from $m = 0.005$ to $m = 1$ concludes that the endpoint of the line of
first-order phase transitions is in the vicinity of $m \approx 0.3$, as seen in
Fig.\ \ref{fig:rhmc_l44_pbp}. The error bars are negligible, outside of rare
cases near the transitions for higher masses, and thus have been omitted from
Fig.\ \ref{fig:rhmc_l44_pbp} to provide greater clarity. A typical example of
the size of the error bars can be seen in the left panel of Fig.\
\ref{fig:rhmc_l44_errors-trans} for $m = 0.05$.

The scaling of these observables with the size of the lattice can be measured,
allowing us to explore the progression of a finite temperature transition into a
bulk transition. We find that the transitions in the chiral condensate $\langle
\overline{\psi} \psi \rangle$ and average plaquette are bulk transitions for
isotropic lattices $V \equiv L_x^3 \times L_t = L^4$ with $L \ge 12$, as seen in
Fig.\ \ref{fig:rhmc_m0,02}. This rapid convergence persists even with improved
actions \cite{Cheng:2012prd}. These qualities make small-lattice experiments
attractive, insofar as they serve to guide the more accurate large-lattice
experiments that are currently in progress.

\begin{figure}
    \begin{center}
    \includegraphics[width=0.60\textwidth]{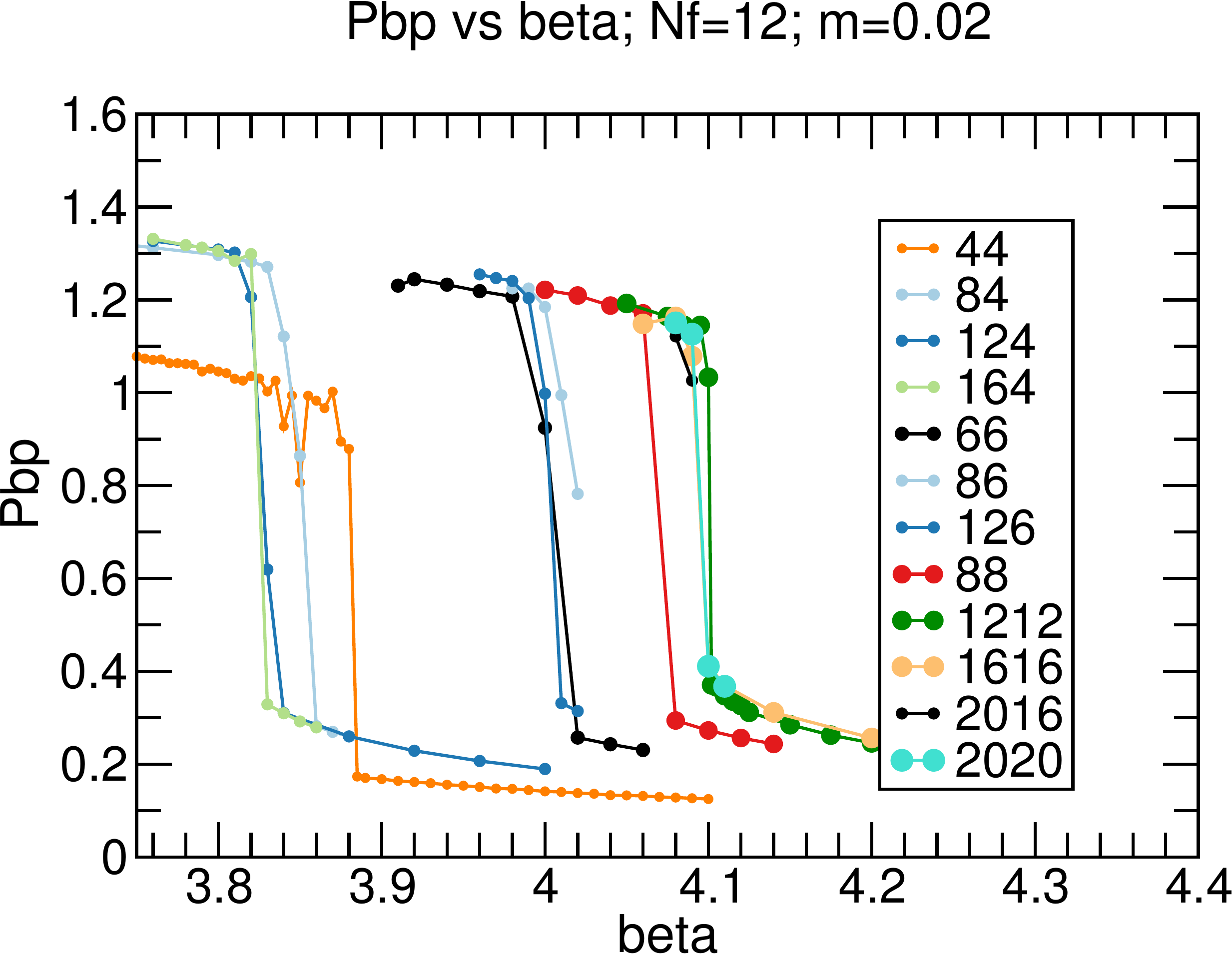}
    \end{center}
    \begin{center}
    \includegraphics[width=0.60\textwidth]{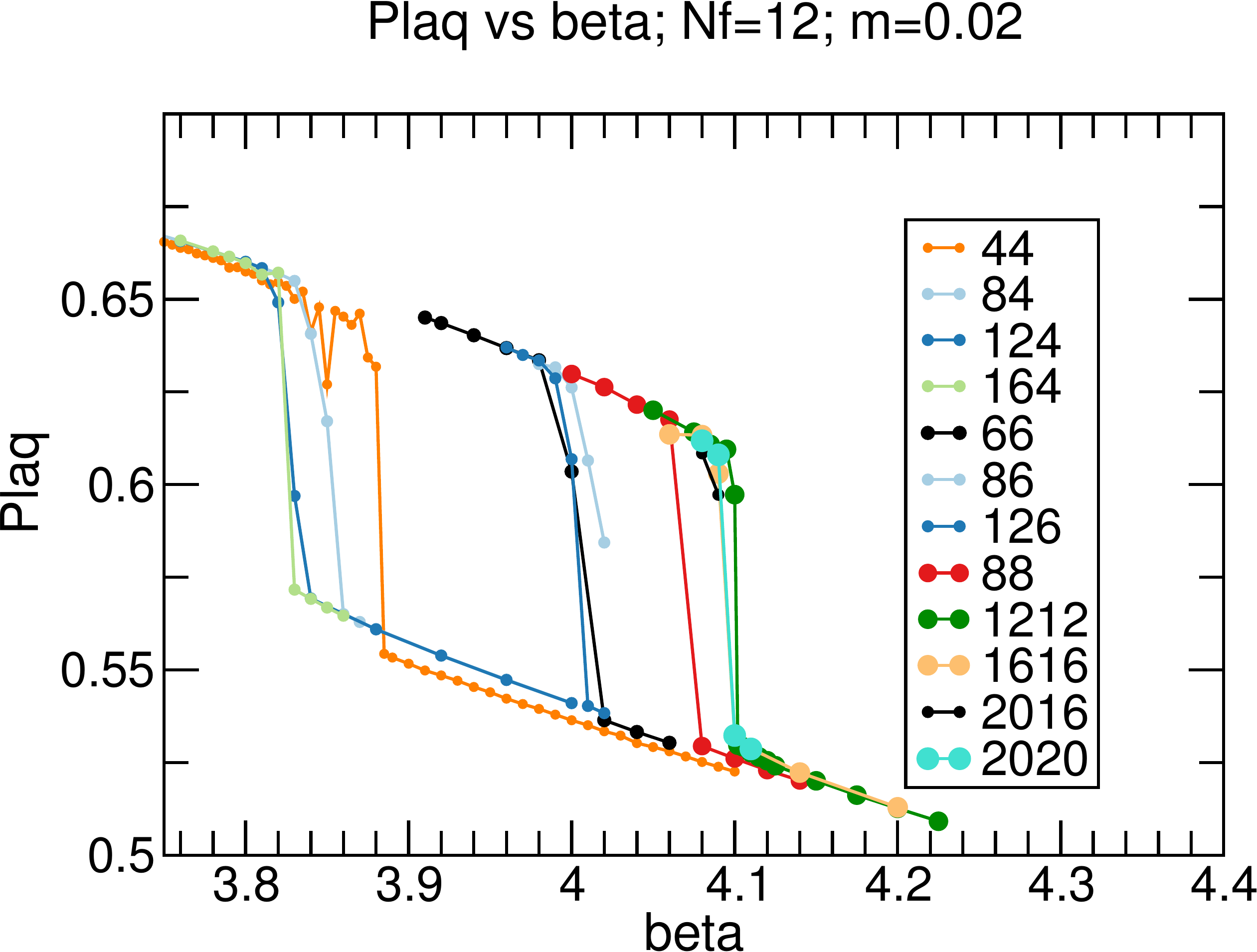}
    \caption{Unimproved HMC.
             Top panel: chiral condensate vs.\
             $\beta$, with $N_f = 12, ~m=0.02$.
             Bottom panel: plaquette vs.\ $\beta$,
             with $N_f = 12, ~m=0.02$.
             In the legend, ``124'' signifies a volume of $V = 12^3 \times 4$.}
    \label{fig:rhmc_m0,02}
    \end{center}
\end{figure}

\begin{figure}
    \begin{center}
    \includegraphics[width=0.60\textwidth]{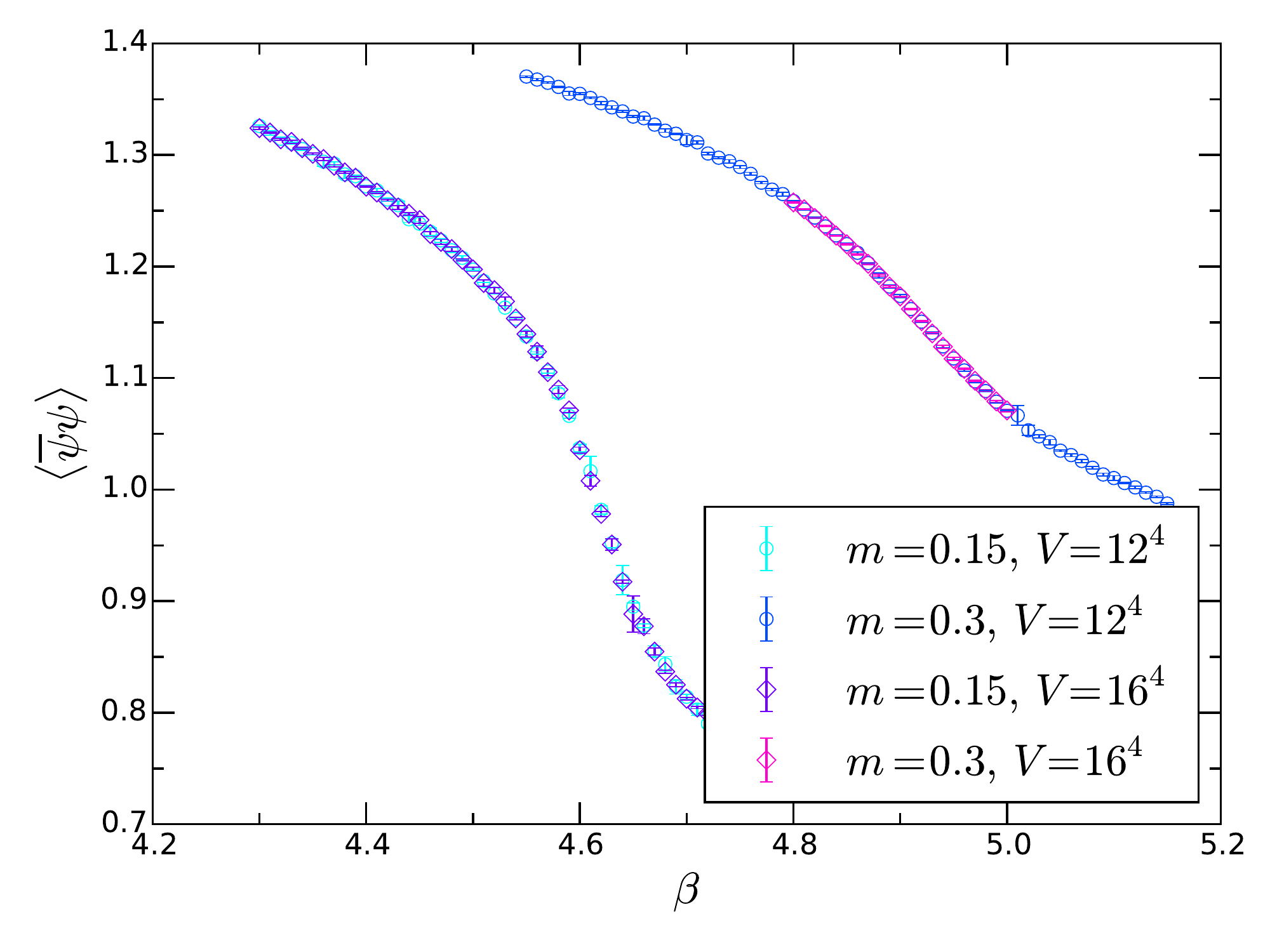}
    \caption{Unimproved HMC.
             Chiral condensate vs.\ $\beta$ for different $m$ and $V$, with $N_f
             = 12$. This shows nearly exact agreement between the two volumes.}
    \label{fig:rhmc_pbp}
    \end{center}
\end{figure}

We are working to find the transition endpoint on larger lattices; results in
Fig.\ \ref{fig:rhmc_pbp} reveal that $m=0.15, 0.30$ are beyond this endpoint.
Preliminary results for $m = 0.075, 0.125$ suggest that the endpoint lies
somewhere in this range. Although the volume effects for $N_f = 12$ unimproved
HMC are negligible beyond $V = 12^4$, we will confirm our proximity to the
continuum with select simulations on $V = 24^4$.

In addition to unimproved HMC, we make use of improved actions. We employ a
gauge action with both fundamental and adjoint plaquette terms, which are tuned
to $\beta_A = -0.25 \beta_F$ to remove a spurious ultraviolet fixed point
(UVFP), a well-known lattice artifact \cite{Cheng:2012prd}. We also employ
smeared staggered fermions. The smearing procedure \cite{Hasenfratz:2007jhep}
aims to alleviate so-called taste splitting: i.e., the symmetry breaking at
nonzero lattice spacing $a$ between the four fermion ``tastes'' described by
each unrooted staggered fermion. We replace the ``thin'' links $U_{n_, \mu}$
(at the lattice site $n$, in the direction $\mu$) by hypercubic (HYP) links that
are defined according to three smearing parameters $\alpha_i$, which are set to
$\alpha_i = (0.75, 0.6, 0.4)$ in Ref.\ \cite{Hasenfratz:2007jhep}. We may then
form normalized HYP (nHYP) links by projecting them to $U(3)$, rather than
$SU(3)$, such that they become differentiable. In keeping with
\cite{Cheng:2012prd}, we adjust to $\alpha_i = (0.5, 0.5, 0.4)$, which improves
the $U(3)$ projection at strong coupling while marginally increasing the taste
splitting.

Applying the smeared links with $N_f = 8, 12$ leads to a novel phase: the broken
shift symmetry $\slashed{S}^4$ phase \cite{Cheng:2012prd}. The single-site shift
symmetry $S^4$ is an exact symmetry of the staggered fermion action, which
ensures that $\langle \overline{\psi} \psi \rangle$ measured on even lattice
sites is identical to that measured on odd ones. Order parameters sensitive to
the spontaneous breaking of $S^4$ include the difference between neighboring
plaquettes $\square_n$,
\begin{equation}
    \Delta P_{\mu} = \left\langle \operatorname{Re} \operatorname{Tr} \square_n
                     - \operatorname{Re} \operatorname{Tr} \square_{n + \mu}
                     \right\rangle_{n_\mu {\rm even}} ,
    \label{eq:DP_mu}
\end{equation}
which becomes nonzero in direction(s) $\mu$ during the $\slashed{S}^4$ phase
\cite{Cheng:2012prd}. A confirmation of the trend of the $\slashed{S}^4$ phase
has been completed for $N_f = 8$; an in-depth look at $N_f = 12$ is ongoing.

\begin{figure}
    \begin{center}
    \includegraphics[width=0.60\textwidth]{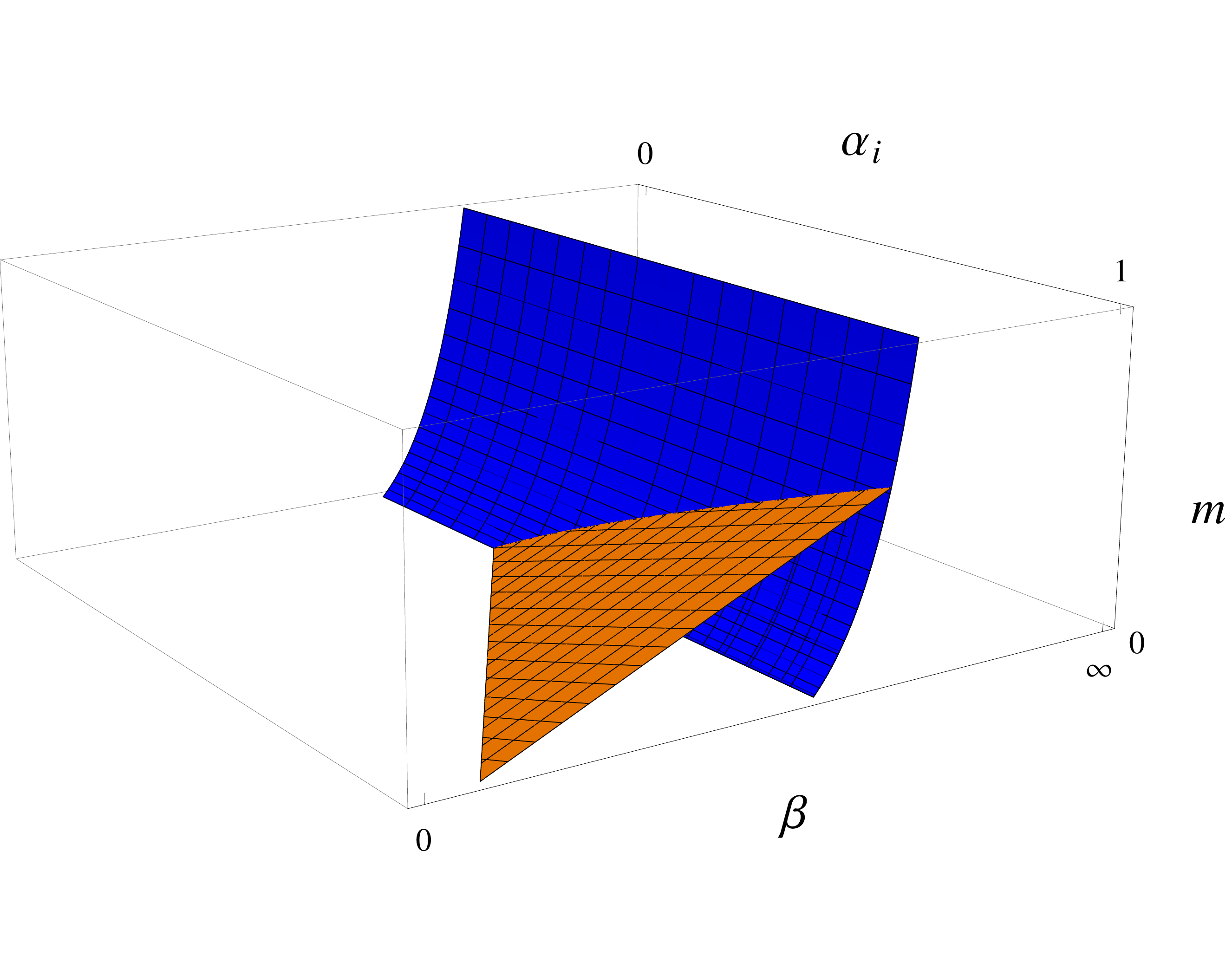}
    \caption{Sketch for nHYP-improved MILC.
             Expected phase diagram of transitions in the
             $m$--$\beta$--$\alpha_i$ space. The orange plane highlights the
             $\slashed{S}^4$ phase, while the blue plane depicts the bulk
             transition between $\chi$SB and deconfinement.}
    \label{fig:nHYP_tricritical}
    \end{center}
\end{figure}

Interpolating between the improved and unimproved classes of actions, we expect
that the $\slashed{S}^4$ phase should vanish as we slowly lessen the improvement
parameters. Simulations are currently in progress to ascertain this effect; a
qualitative shape of the expected phase diagram is included in Fig.\
\ref{fig:nHYP_tricritical}.

\section{Conclusions}

For the unimproved class of simulations, we will continue to seek the endpoint
of the line of first-order phase transitions, which we currently believe exists
between $m = 0.075$ and $m = 0.125$. Once at this endpoint, we will conduct an
analysis with Fisher zeros around the critical beta $\beta_C$, as discussed
by Y.\ Liu et al.\ in Ref.\ \cite{Liu:2013posl}.

For the improved class of simulations, we will seek to confirm the
$\slashed{S}^4$ phase for $N_f = 12$. Given that $S^4$ is expected to be
reinstated in some region of the improvement parameter space $\alpha_i,
\beta_A$, interesting phenomena may appear at the boundary of this region. A
tricritical point may exist between the boundaries of the phases of $\chi$SB
(chiral symmetry breaking), $\slashed{S}^4$, and deconfinement (chiral symmetry
restoring).

\subsection*{Acknowledgements}

We thank Dr.\ Donald Sinclair of Argonne National Laboratory for the use of his
HMC code.

We also thank Dr.\ Anna Hasenfratz of the University of Colorado, Boulder, for
the assistance she has provided in the tuning of the improvement parameters of
the nHYP-improved MILC code.

Finally, we thank MILC for continuing to make its code readily available under
the GNU General Public License.

This research was supported in part by the Department of Energy under Award
Numbers DE-SC0010114 and FG02-91ER40664. We used the National Energy Research
Scientific Computing Center (NERSC), which is supported by the Office of Science
of the U.S.\ Department of Energy under Contract No.\ DE-AC02-05CH11231. Z.\ G.\
is supported by the URA Visiting Scholars program and uses the Fermilab Lattice
Gauge Theory Computational Facility. Fermilab is operated by Fermi Research
Alliance, LLC, under Contract No.\ DE-AC02-07CH11359 with the U.S.\ Department
of Energy.

\end{document}